%% file: main.tex
\documentclass[a4paper,12pt,britishm,texlive=2016]{article}

\RequirePackage{graphicx}
\RequirePackage{subcaption}
\RequirePackage{siunitx}
\RequirePackage{mhchem}
\RequirePackage{url}
\RequirePackage{xcolor}
\RequirePackage{bbold}
\RequirePackage{placeins}
\RequirePackage[numbers,sort&compress]{natbib} 

\RequirePackage[colorlinks=true, allcolors=blue]{hyperref}

\newcommand{\TeV}{\ensuremath{\text{Te\kern -0.1em V}}}
\newcommand{\GeV}{\ensuremath{\text{Ge\kern -0.1em V}}}
\newcommand{\MeV}{\ensuremath{\text{Me\kern -0.1em V}}}
\newcommand{\pt}{\ensuremath{p_{\text{T}}}}
\newcommand{\met}{\ensuremath{E_{\text{T}}^{\text{miss}}}}
\newcommand{\tH}{\ensuremath{tH}}
\newcommand{\ttH}{\ensuremath{t\bar{t}H}}
\newcommand{\Hgg}{\ensuremath{H\to\gamma\gamma}}

\makeatletter
\g@addto@macro\bfseries\boldmath
\makeatother

\title{KAN we improve on HEP classification tasks? \\ Kolmogorov-Arnold Networks \\ applied to an LHC physics example}

\author{Johannes Erdmann, Florian Mausolf, Jan Lukas Sp{\"a}h}

\date{\small
  RWTH Aachen University, III. Physikalisches Institut A, Aachen, Germany
  }

\begin{document}

\maketitle

\begin{abstract}
  Recently, Kolmogorov-Arnold Networks (KANs) have been proposed as an alternative to multilayer perceptrons, suggesting advantages in performance and interpretability.
  We study a typical binary event classification task in high-energy physics including high-level features and comment on the performance and interpretability of KANs in this context.
  Consistent with expectations, we find that the learned activation functions of a one-layer KAN resemble the univariate log-likelihood ratios of the respective input features.
  In deeper KANs, the activations in the first layer differ from those in the one-layer KAN, which indicates that the deeper KANs learn more complex representations of the data, a pattern commonly observed in other deep-learning architectures.
  We study KANs with different depths and widths and we compare them to multilayer perceptrons in terms of performance and number of trainable parameters.
  For the chosen classification task, we do not find that KANs are more parameter efficient.
  However, small KANs may offer advantages in terms of interpretability that come at the cost of only a moderate loss in performance.
\end{abstract}

\clearpage

\section{Introduction}
\label{sec:introduction}

\input{tex_inputs/introduction.tex}

\section{Kolmogorov-Arnold Networks}
\label{sec:methods}

\input{tex_inputs/ML_methods.tex}

\section{Dataset}
\label{sec:simulation}

\input{tex_inputs/samples.tex}

\FloatBarrier

\section{Results}
\label{sec:results}
\input{tex_inputs/results.tex}

\FloatBarrier

\section{Conclusions}
\label{sec:conclusions}

\input{tex_inputs/conclusions.tex}

\input{tex_inputs/acknowledgements.tex}

\bibliographystyle{JHEP}
\bibliography{main}

\section*{Appendix}
\label{sec:appendix}

\input{tex_inputs/appendix.tex}

\end{document}

%% file: tex_inputs/introduction.tex
Classifying events as signal or background is a crucial ingredient of data analysis at collider experiments.
At the Large Hadron Collider (LHC), separating small signals from large backgrounds is an omnipresent challenge.
To achieve higher precision in the analysis of collider data, excellent classifiers are necessary.
Machine-learning-based classifiers have a long history in high-energy physics (HEP).
For example, the observation of electroweak production of single top quarks in 2009 at the Tevatron~\cite{D0:2009isq,CDF:2009itk} was aided by boosted decision trees and by shallow neural networks, i.e., multilayer perceptrons (MLPs) with one hidden layer.
With the development of deep neural networks, MLPs with several hidden layers have been proposed for HEP classification tasks~\cite{Baldi:2014kfa} and have become a standard tool for event classification, particle identification, fast simulations and many more applications at the LHC~\cite{Feickert:2021ajf,Guest:2018yhq,Radovic:2018dip,Schwartz:2021ftp,Karagiorgi:2021ngt}.

The strong performance of MLPs comes with a trade-off in terms of interpretability.
Interpretability, i.e., the ``ability to explain or to present in understandable terms to a human''~\cite{interpret_def}, remains a challenge for MLPs, particularly for deep networks with many trainable parameters.
At the same time, understanding what such a model has learned about the underlying physics is of genuine interest in physics applications.
Several methods have been developed to address this challenge of explaining the outputs of MLPs for given input examples~\cite{XAI, guidotti2018survey, interpretable}.
Techniques such as Shapley values~\cite{SHAP} and permutation feature importance~\cite{Breiman2001} are established methods for assessing the contribution of individual input features to the model output.
Surrogate models, such as LIME~\cite{LIME}, aim to explain the reasoning of complex architectures by approximating them with simplified models.
On the other hand, approaches like Neural Additive Models~\cite{NAMs} provide interpretability by constructing models that are transparent by design.

Recently, Kolmogorov-Arnold Networks (KANs) have been proposed as an alternative to MLPs~\cite{KAN}.
While MLPs are grounded in the universal approximation theorem~\cite{UAT}, KANs are motivated by the Kolmogorov-Arnold representation theorem~\cite{kolmogorov}.
The layers of the KAN have learnable activation functions on the edges that are summed on the nodes.
In contrast, MLP layers use learnable weights on the edges as inputs to fixed activation functions on the nodes.
Many approaches have been explored to improve the expressiveness and performance of MLPs by introducing learnable activation functions.
Examples include implementations based on splines \cite{Bohra, Aziznejad}, other parametric functions \cite{agostinelli2014learning, goyal2019learning, SNAKE, DRA} or even neural networks \cite{zhang2022neural} as activation functions.

While networks based on the Kolmogorov-Arnold representation theorem were proposed before~\cite{sprecher,koeppen,lin,lai,leni,fakhoury,montanelli,he}, recently the capabilities of KANs in terms of performance and interpretability were highlighted~\cite{KAN}.
In Ref.~\cite{KAN}, KANs were found to have promising performance with a substantially smaller number of trainable parameters than MLPs.
KANs offer advantages in terms of interpretability, complementarily to existing explainability methods applicable to both KANs and MLPs, due to their shallower structures with significantly fewer nodes than typical MLPs.
Each edge in a KAN contains multiple trainable parameters that determine the shape of a single function.
Therefore, an entire KAN can be represented as a comprehensive graph.
In addition, the potential for interpretability by approximating the learned activation functions symbolically with a set of known functions was discussed.
Ref.~\cite{KAN} has sparked active discussion on the potential advantages of KANs and their relation to MLPs~\cite{duda2024biology, li2024kolmogorov, genet2024tkan, peng2024predictive, vaca2024kolmogorov, samadi2024smooth, bozorgasl2024wav, yang2024endowing, abueidda2024deepokan, cheon2024kolmogorov, xu2024fourierkan, xu2024kolmogorov, genet2024temporal, nehma2024leveraging, li2024u, shukla2024comprehensive, herbozo2024kan, kiamari2024gkan, aghaei2024fkan, seydi2024unveiling, azam2024suitability, chen2024sckansformer, wang2024kolmogorov, ta2024bsrbf, zhang2024graphkan, bodner2024convolutional, poeta2024benchmarking, aghaei2024rkan, cheon2024demonstrating, de2024kolmogorov, bresson2024kagnns, howard2024finite, wang2024spectralkan, lobanov2024hyperkan, dong2024tckin, lawan2024mambaforgcn, altarabichi2024dropkan, shen2024reduced, inzirillo2024deep, troy2024sparks, toscano2024inferring, li2024coeff, rigas2024adaptive, le2024exploring, seguelvlp, zeydan2024f, zinage2024dkl, pratyush2024calmphoskan, altarabichi2024rethinking, liu2024complexity, tang20243d, li2024gnnmolkanharnessingpowerkan}.

We apply KANs to a typical HEP event classification task.
As an example, we choose the binary separation of the associated production of a Higgs boson with a single top quark (\tH) and with a top quark and an anti-top quark (\ttH) at the LHC, where the Higgs boson decays to a pair of photons (\Hgg).
We study the interpretability of KANs for this classification task.
In addition, we compare KANs to MLPs in terms of performance and parameter efficiency, where we use KANs and MLPs with different numbers of layers and nodes per layer.
We document our findings in the practical training of KANs.
To our knowledge, this is the first application of KANs to a task in particle physics.

%% file: tex_inputs/ML_methods.tex
For the comparison to KANs, we briefly summarize the concept of MLPs.
An MLP consists of multiple layers of nodes, each connected to nodes in subsequent layers through weighted edges.
The core component of an MLP is the fully connected layer, which holds the trainable parameters defining the strength of the connections between nodes of two layers.
Each layer applies an affine transformation, represented by a weight matrix $\mathbf{W}$ and a bias vector $\vec{b}$, followed by an activation function $\mathcal{A}$.
The transformation applied in each MLP layer can then be written as $\vec{y} = \mathcal{A}\left(\mathbf{W} \vec{x} + \vec{b}\right)$, where $\vec{x}$ denotes the input to the layer and $\vec{y}$ is its output.
The activation function introduces non-linearity in the model and is a hyperparameter that has to be chosen.
Common choices include the rectified linear unit $\mathrm{ReLU}(x)=\max{(0,x)}$, the logistic sigmoid function $\sigma(x)$, and the hyperbolic tangent function.

In contrast, KANs are inspired by the Kolmogorov-Arnold representation theorem, which states that any continuous multivariate function $f: [0, 1]^n \to \mathbb{R}$ can be represented as a finite sum of continuous functions of only one variable.
Formally, for any continuous real-valued function $f(x_1, x_2, \ldots, x_n)$, continuous functions $\phi_{i(j)}$ exist, such that
\begin{equation}
    f(x_1, x_2, \ldots, x_n) = \sum_{i=1}^{2n+1} \phi_i \left( \sum_{j=1}^n \phi_{ij}(x_j) \right),
    \label{eq:KAT}
\end{equation}
where $n$ is the number of variables that parameterize the multivariate function, and $\phi_i$ and $\phi_{ij}$ are univariate functions.
This representation reduces the problem of approximating a multivariate function to a problem involving only univariate functions and the sum operation.
The objective of the network training is to approximate these univariate functions $\phi_{i(j)}$.

Motivated by the theorem, the appropriate network architecture to approximate a multivariate function of $n$ variables consists of two layers with $n$ input nodes, $2n+1$ hidden nodes, and a single output node.
However, the authors of Ref.~\cite{KAN} generalized the concept by defining a KAN layer as a basic building block.
As in MLPs, the number of nodes in these layers can be customized and layers can be stacked arbitrarily to enhance the performance of the model.
Similar to MLPs, each node in a given layer is connected to each node of the subsequent layer.
For each edge, an individual, learnable activation function is used.
On the nodes, only the sum operation over all incoming edges is performed.

In the implementation of Ref.~\cite{KAN}, the learnable activation functions are defined as the weighted sum of a B-spline, expressed by B-spline basis functions $B_i$, and a fixed residual function, chosen as the sigmoid-linear unit $\mathrm{SiLU}(x) = x\cdot \sigma(x)$:
\begin{equation}
    \mathrm{activation}(x) = w_1\cdot \mathrm{SiLU}(x) + w_2\cdot \sum_{i=0}^{G+k-1}c_i\cdot B_i(x).
\end{equation}

The weights $w_{1,2}$ and the basis-function coefficients $c_i$ are the trainable parameters of the spline.
The basis functions $B_i$ are chosen as polynomials of degree $k$, with default value $k=3$.
The grid parameter $G$ determines how many basis functions build the B-spline and serves as a hyperparameter of the KANs.
Furthermore, the domain of the activation function needs to be chosen and can be updated several times during network training to match the input range of the activation function. 
Specifically, for given parameters $k$, $G$ and the domain $[t_0, t_G]$, a vector $\vec{t} = (t_{-k}, \ldots, t_0, \ldots, t_{G+k})$ of equidistant knot points is constructed.
Then, $G+k$ basis functions $B_i^k(x)$ are recursively defined:
\begin{equation}
B_i^0(x) = 
\begin{cases} 
    1 & \text{if } t_i \leq x < t_{i+1}, \\
    0 & \text{otherwise},
\end{cases}
\end{equation}
and for $k > 0$:
\begin{equation}
    B_i^k(x) = \frac{x - t_i}{t_{i+k} - t_i} B_i^{k-1}(x) + \frac{t_{i+k+1} - x}{t_{i+k+1} - t_{i+1}} B_{i+1}^{k-1}(x).
\end{equation}

The basis functions are only non-zero over a portion of the interval, allowing the coefficients $c_i$ to adapt and change the overall spline locally.
This enables the approximation of functions without strong assumptions about their functional form.
More details on the implementation can be found in Ref.~\cite{KAN}.

%% file: tex_inputs/samples.tex
As an example for a typical HEP classification task, we use the separation of \ttH\ from \tH\ production in the \Hgg\ decay channel at the LHC.
Both processes offer complementary sensitivity to properties of the Yukawa coupling of the top quark~\cite{Farina:2012xp,Bahl:2020wee}, but only \ttH\ production has been observed so far~\cite{CMS:2018uxb,ATLAS:2018mme}.
In the search for \tH\ production, the \Hgg\ decay offers one of the most sensitive channels~\cite{CMS:2024fdo,ATLAS:2022vkf}, for which \ttH\ is a major background and hence excellent binary classification is necessary.

We simulate \ttH\ and \tH\ production\footnote{For \tH\ production, the charge-conjugate process is also included, but we denote the sum of both processes as \tH\ for simplicity.} in proton-proton collisions at a center-of-mass energy of \mbox{$14\,\TeV$}.
We use \texttt{MadGraph5\_aMC@NLO}~\cite{AMCAT} (version 3.5.3) at leading order in perturbative quantum chromodynamics for the hard-scattering processes with the \texttt{NNPDF23\_lo\_as\_0130\_qed}~\cite{Ball:2012cx} set of parton distribution functions.
We use the five-flavor scheme for the simulation of \ttH\ production.
The four-flavor scheme is chosen for \tH production for an improved event modeling~\cite{Demartin:2015uha}, where only the dominant $t$-channel contribution is considered.
Only events with at least one semi-leptonic top quark decay are simulated\footnote{Final states with $\tau$ leptons are included in the event generation.}.
For both processes, the factorization and renormalization scales are set event-by-event to the transverse mass of the irreducible $2\to2$ system resulting from a $k_{\mathrm{T}}$ clustering of the partons in the final state~\cite{loop_induced}.
The events are interfaced to Pythia~8.3.1~\cite{Bierlich:2022pfr} for the \Hgg\ decay, parton shower and hadronization.
We use Delphes~3.5.1~\cite{deFavereau:2013fsa} for a fast simulation of the CMS detector response with the CMS card with default settings.
These settings include jet clustering with the anti-$k_T$ algorithm~\cite{Cacciari:2008gp, Cacciari:2011ma} with a radius parameter of $R=0.5$. 

We focus on final states with two photons, at least one charged lepton (electron or muon), at least one $b$-jet and at least one additional jet.
The following requirements are applied, where $\pt$ is the transverse momentum and $\eta$ is the pseudorapidity:
\begin{itemize}
  \item exactly two photons (ordered in \pt) with $\pt(\gamma_1) > 35 \,\GeV$ and \\$\pt(\gamma_2) > 25 \,\GeV$ and $|\eta(\gamma_{1,2})| < 2.5$;
  \item invariant diphoton mass in the range $100\,\GeV < m(\gamma\gamma) < 180\,\GeV$ with $\pt(\gamma_1) / m(\gamma\gamma) > \genfrac{}{}{}{1}{1}{3}$ and $\pt(\gamma_2) / m(\gamma\gamma) > \genfrac{}{}{}{1}{1}{4}$;
  \item at least one charged lepton with $\pt(\ell) > 10\,\GeV$ and $|\eta(\ell)| < 2.4$;
  \item at least one $b$-jet with $\pt(b) > 25\,\GeV$ and $|\eta(b)| < 2.5$;
  \item at least one additional jet with $\pt(j) > 25\,\GeV$ and $|\eta(j)| < 4.7$.
\end{itemize}
After applying these selection criteria, we have \num{100000} events for training the classifiers and \num{33000} events for validation during training, for each of the two processes.
To ensure small statistical uncertainties in the metrics used in this study, we use \num{100000} events per process of an independent test set to evaluate the networks.

\begin{figure}[t]
    \centering
    \includegraphics[width=\textwidth]{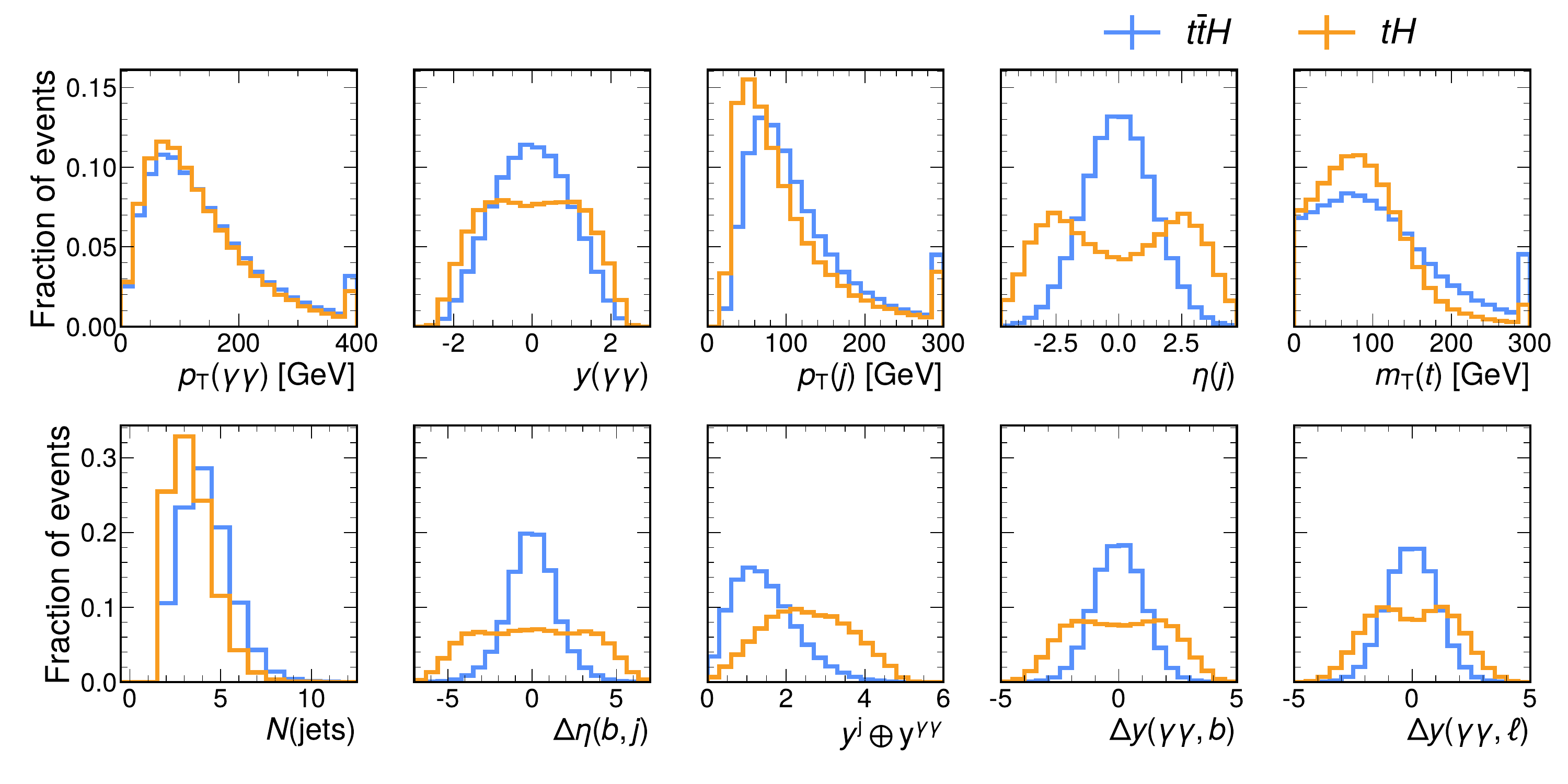}
    \caption{Distributions of ten example features used for the classification.
    For distributions with overflow, the overflow is included in the last bin.
    }
    \label{fig:inputFeatures}
\end{figure}

We use 22 input features, which include four-vector components and high-level features based on photons, charged leptons, the missing transverse momentum (\met) and jets\footnote{While most of the high-level features are standard observables, the variable $y^j \oplus y^{\gamma\gamma}$ was proposed in Ref.~\cite{Bahl:2020wee} to disentangle \ttH\ and \tH\ production when testing different CP hypotheses for the top-quark Yukawa coupling.}.
Among the selected jets, we focus on the highest-\pt\ $b$-jet and the leading jet of the event excluding this $b$-jet (``additional jet'').
Ten of the features are shown in Fig.~\ref{fig:inputFeatures} for the two event classes.
As it is typical for HEP event classification, no single feature provides sufficient discrimination on its own.
Several features show the expected differences between \ttH\ and \tH\ production.
For example, the number of jets ($N(\mathrm{jets})$) is larger in \ttH\ production given the second top quark in the final state, and the pseudorapidity of the additional jet ($\eta(j)$) tends towards larger absolute values for \tH\ due to the electroweak $t$-channel topology.

\begin{figure}[t]
    \centering
    \includegraphics[width=0.85\textwidth]{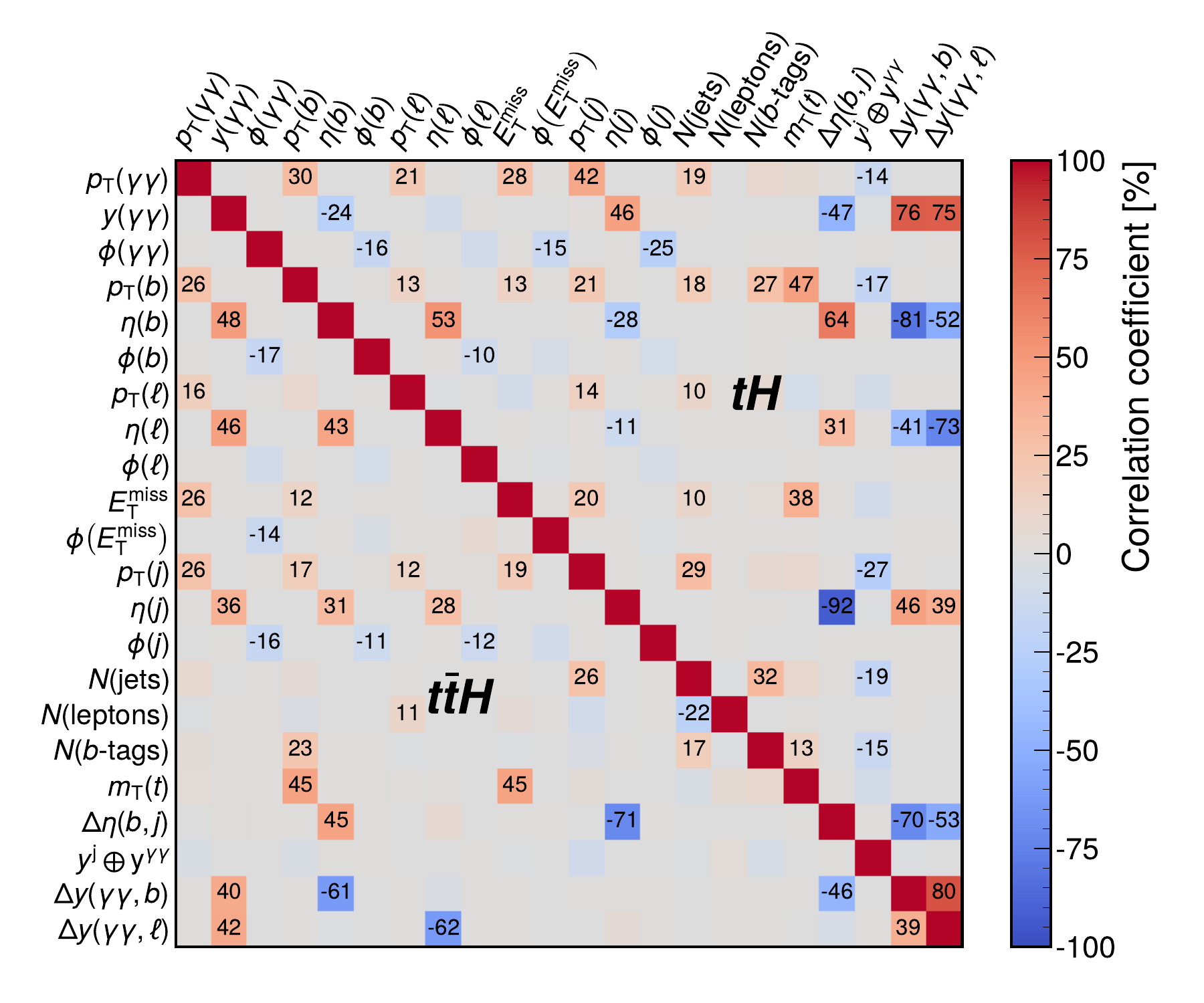}
    \caption{Matrix of the Pearson correlation coefficients of all 22 input features.
    The upper triangle refers to the \tH\ dataset and the lower triangle refers to the \ttH\ dataset.
    Off-diagonal coefficients with absolute values of at least $10\,\%$ are shown as numbers on the plot.
    }
    \label{fig:correlationMatrix}
\end{figure}

The matrix of the Pearson correlation coefficients is shown in Fig.~\ref{fig:correlationMatrix}, separately for \ttH\ (lower triangle) and \tH\ production (upper triangle).
The 22 features show a non-trivial correlation structure with strong positive and negative correlations between some of the features.
The correlations differ significantly in the \ttH\ and \tH\ datasets.
For example, while the distributions of the transverse momentum of the diphoton system ($\pt(\gamma\gamma)$) in Fig.~\ref{fig:inputFeatures} are almost identical for the two datasets, the correlations of $\pt(\gamma\gamma)$ with other features are not.

%% file: tex_inputs/results.tex
We compare KANs and MLPs of different configurations for the classification of \ttH\ and \tH\ events.
The KANs are implemented using the \texttt{Pykan} package from Ref.~\cite{KAN} in version 0.2.1 together with \texttt{PyTorch}~\cite{paszke2019pytorch} version 2.3.0.
All MLPs are implemented with \texttt{TensorFlow}~2.17.0~\cite{tensorflow_2_17_0}.

We scale all input features before feeding them to the networks.
A common approach for MLPs is studentization, where the sample mean is subtracted from each variable and the values are divided by the sample standard deviation.
We apply this method for all MLP trainings.
For KANs, we apply a different transformation to avoid the impact of outliers far away from the bulk of the distributions, which are common in typical HEP datasets.
KANs require the domain of each learnable activation function to be defined by the range of the input for each spline.
Outliers can extend the domain boundaries beyond the bulk of the distributions.
As a result, the spline that acts on the majority of events may be parametrized by only a small fraction of the basis functions, which reduces the flexibility of curve approximation for the most relevant domain.
To mitigate this, we first apply a logarithmic transformation to all transverse momenta and the transverse top quark mass, $\tilde{x}= \ln\left(1 + x\right)$, where $x$ denotes the observable in units of \GeV.
We then apply min-max scaling in the range $[0,1]$ and initialize the spline domains accordingly.

The output layers of all trained models consist of a single node.
Although KANs learn activation functions and a learnable function can also be placed on the output node, we choose the sigmoid function to normalize the model outputs to the range $(0, 1)$.
In addition, for the MLPs, we choose the sigmoid function as output activation.
For all trainings, we use the binary cross-entropy as a loss function.
We consider \tH\ events as signal (label~1) and \ttH\ as background (label~0).

We use the Adam~\cite{kingma2017adam} optimizer to train our models.
For the KAN trainings, we also compare with the Limited-memory Broyden--Fletcher--Goldfarb--Shanno (L-BFGS)~\cite{LBFGS} optimizer, as it was used in Ref.~\cite{KAN}.
For Adam, we find stable trainings for all tested learning rates in the range $[10^{-4}, 10^{-3}]$ and select $3\times 10^{-4}$ for all trainings.
The training is performed in mini-batches of batch size \num{256}.
Trainings with the L-BFGS optimizer are performed using the entire dataset in full-batch training with a learning rate of $10^{-3}$.
No significant performance differences between KANs trained with the two optimizers were found and hence we use Adam for all trainings discussed below due to its faster convergence.
To ensure that all models converge during training and to allow for a fair performance comparison, we employ early stopping.
The loss obtained from the validation dataset is monitored during training and if there is no improvement over 25 epochs, the training is terminated.
The model parameters from the epoch with the lowest loss on the validation set are used for the comparisons.

We compare multiple KAN structures: The first model uses the configuration inspired by the Kolmogorov-Arnold theorem and hence consists of two layers with a node structure of \mbox{22--45--1}.
This is compared to the simplest possible KAN with only a single layer (22--1), as well as other models with varying widths and depths.
For these models, we choose node structures of 22--3--1, 22--10--5--2--1 and 22--45--10--5--2--1.
We use the default grid parameter, $G=5$, and the default degree of the basis functions, $k=3$.

\begin{figure}[t]
    \centering
    \includegraphics[width=0.9\textwidth]{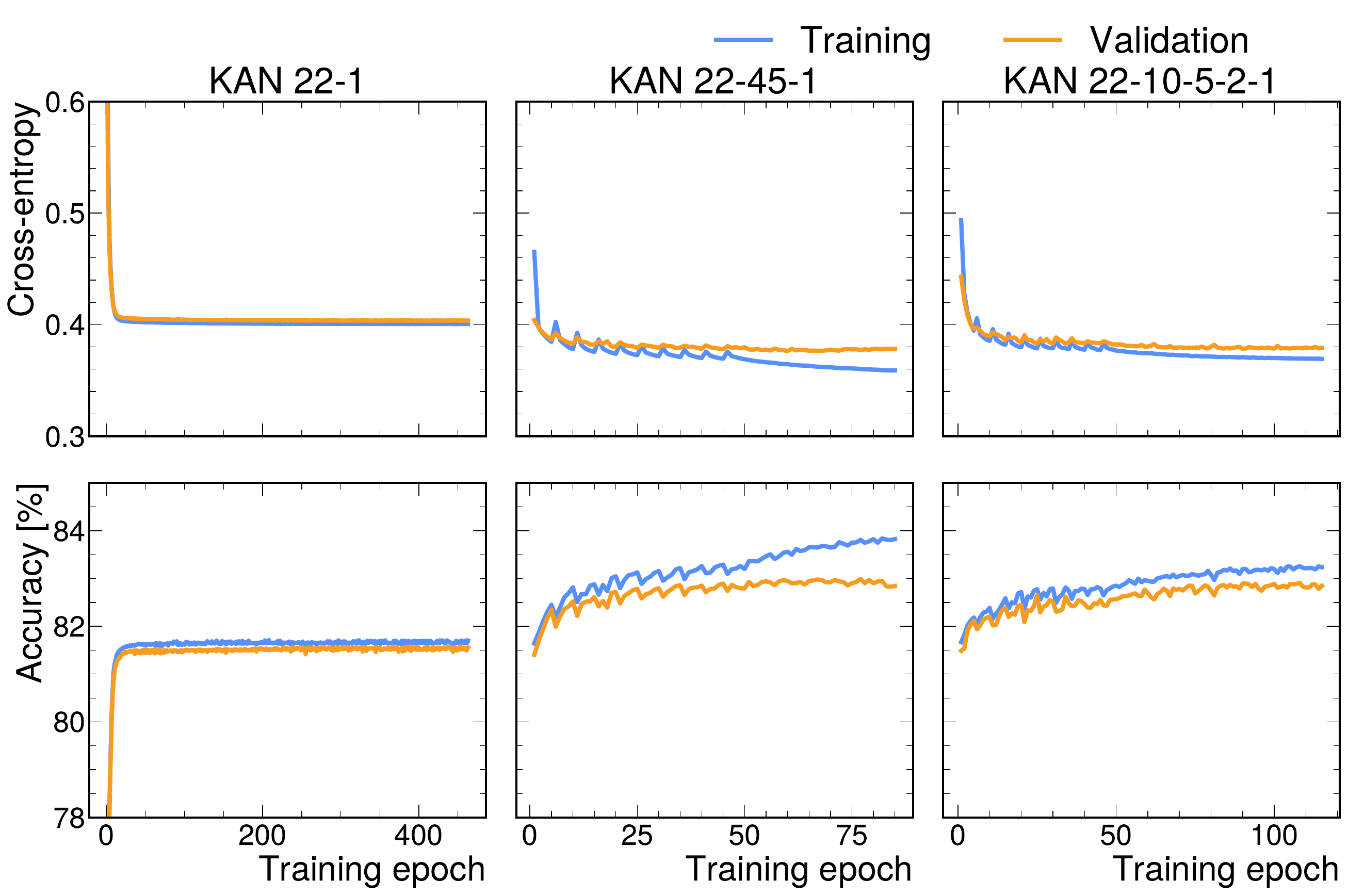}
    \caption{Evolution of the loss (upper row) and the accuracy (lower row) of three KAN models of depth one, two and four, respectively.
    Due to the early-stopping approach, the epoch from which the model parameters are used appears 25 epochs before the end of the optimization.
    Instabilities occur in training epochs where the spline domains of multi-layer KANs are adapted.
    }
    \label{fig:KAN_histories}
\end{figure}

The evolution of the loss and accuracy\footnote{The accuracy is defined as the fraction of correctly classified examples with a decision threshold of 0.5 in the network output.} over the training is shown in Fig.~\ref{fig:KAN_histories} for three KANs.
As expected, the single-layer KAN shows the lowest performance among these models but still reaches an accuracy of $81.5\%$ on the validation set, which is only about $1.5$ percentage points lower than the accuracy of the two-layer and four-layer KANs.
The latter two models reach similar accuracies and loss values.
The two-layer KAN, with its wide second layer, has the highest number of trainable parameters in this comparison.
It converges within the fewest number of training epochs and at the same time shows the largest generalization gap.
While the range of the input variables is fixed with the min-max scaling, this is not the case for the input range of subsequent layers in the networks.
Therefore, we use the default setting to update the domains ten times within the first 50 epochs of training.
Training instabilities are visible at these epochs.

\begin{figure}[t]
    \centering
    \includegraphics[width=0.9\textwidth]{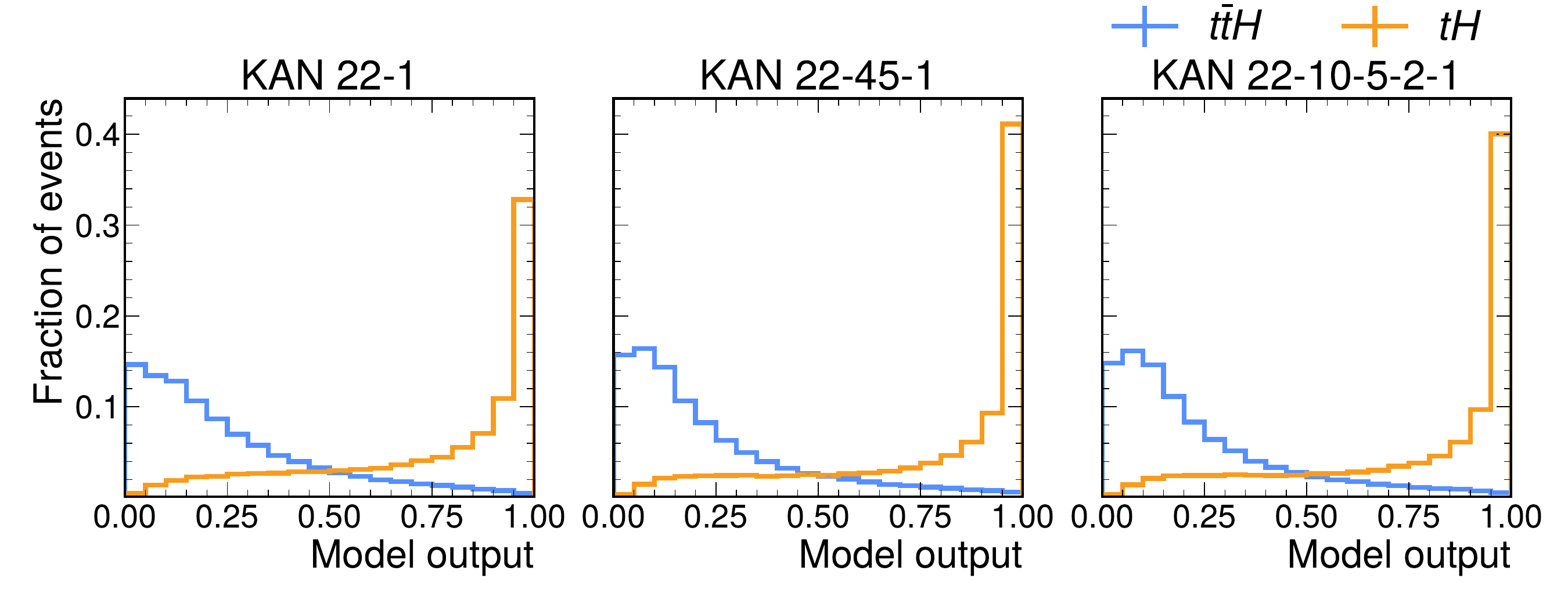}
    \caption{Output distributions on the test dataset for the two classes for three KANs with structures 22--1, 22--45--1 and 22--10--5--2--1, respectively.
    }
    \label{fig:KAN_outputs}
\end{figure}

The output distributions of the test dataset classified by these KANs are shown in Fig.~\ref{fig:KAN_outputs}.
The separation of the two classes is clearly visible with the networks accumulating the majority of the events close to the respective label.
Also here, only slight differences are visible between the two- and the four-layer network, while the one-layer KAN achieves a visibly worse separation.

A possible advantage of KANs over MLPs lies in their potential for interpretability.
While the patterns learned by MLPs are usually embedded in large matrices of trainable parameters and thus hard to understand, multiple trained KAN parameters can be visualized in form of a single spline.
As demonstrated in Ref.~\cite{KAN}, patterns learned by KANs can often be understood more easily.
However, these examples are much lower in dimensionality and complexity than typical HEP machine-learning tasks.
In our study with 22 input features, we find that the interpretability of wide models, such as the 22--45--1 KAN, is limited.
For instance, this particular model consists of more than \num{1000} learnable activation functions.
Its visualization is, hence, complex and difficult for humans to interpret.
Therefore, we focus on shallow KAN structures for the interpretability.

\begin{figure}[t]
    \centering
    \includegraphics[width=\textwidth]{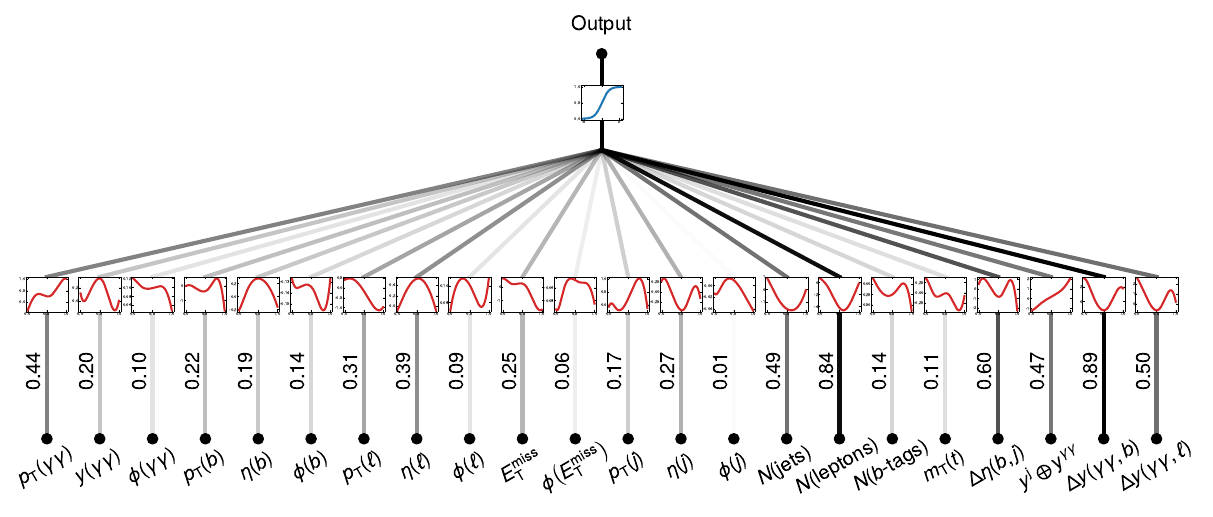}
    \caption{Graphical representation of the trained KAN with a single layer (KAN~22--1).
    The red curves represent the learned activation functions, while the blue curve shows the sigmoid function used to normalize the network output.
    The $L_1$-norm of each spline is given, which also defines the grayscale of each edge.}
    \label{fig:KAN_22-1}
\end{figure}

In Fig.~\ref{fig:KAN_22-1}, the one-layer KAN is depicted together with its learned activation functions.
The corresponding figure for the KAN with a shallow second layer (22--3--1) is shown in the Appendix.
In the following, we discuss the interpretability of the learned activation functions.

The strength of the edges is indicated by the grayscale and it is estimated by the $L_1$-norms of the learned activations, defined as the mean magnitude over the examples as 
\begin{equation}
    \left|\phi(x)\right|_1 = \frac{1}{N(\mathrm{events})}\sum_{i=1}^{N(\mathrm{events})} \left|\phi(x_i)\right|.
\end{equation}
These values directly indicate the importance of the different input features for the one-layer KAN classification output due to the simple summation of input features transformed by a single activation function on its output node.
The feature importances obtained from the commonly used techniques of permutation feature importance and Shapley values yield similar patterns.
This confirms that the $L_1$-norms in the one-layer KAN effectively capture feature importance.
This network is expected to exploit the features with strong discrimination between the two classes.
Values of the $L_1$-norms close to zero are found for the azimuthal angles, which provide no discrimination power on their own.
The largest values of the $L_1$-norms are found for the lepton multiplicity, with which the network can identify di-leptonic signatures present in a fraction of the \ttH\ events but mostly absent in the \tH\ process, and for variables based on (pseudo-)rapidity differences as well as the jet multiplicity.
These reflect the good separation that can be achieved with these variables on their own.
In deeper KANs, where multiple learnable functions transform each feature and their outputs are combined in a more complex manner, the $L_1$-norms are not directly interpretable as feature importance scores.

\begin{figure}[p]
    \centering
    \includegraphics[width=0.8\textwidth]{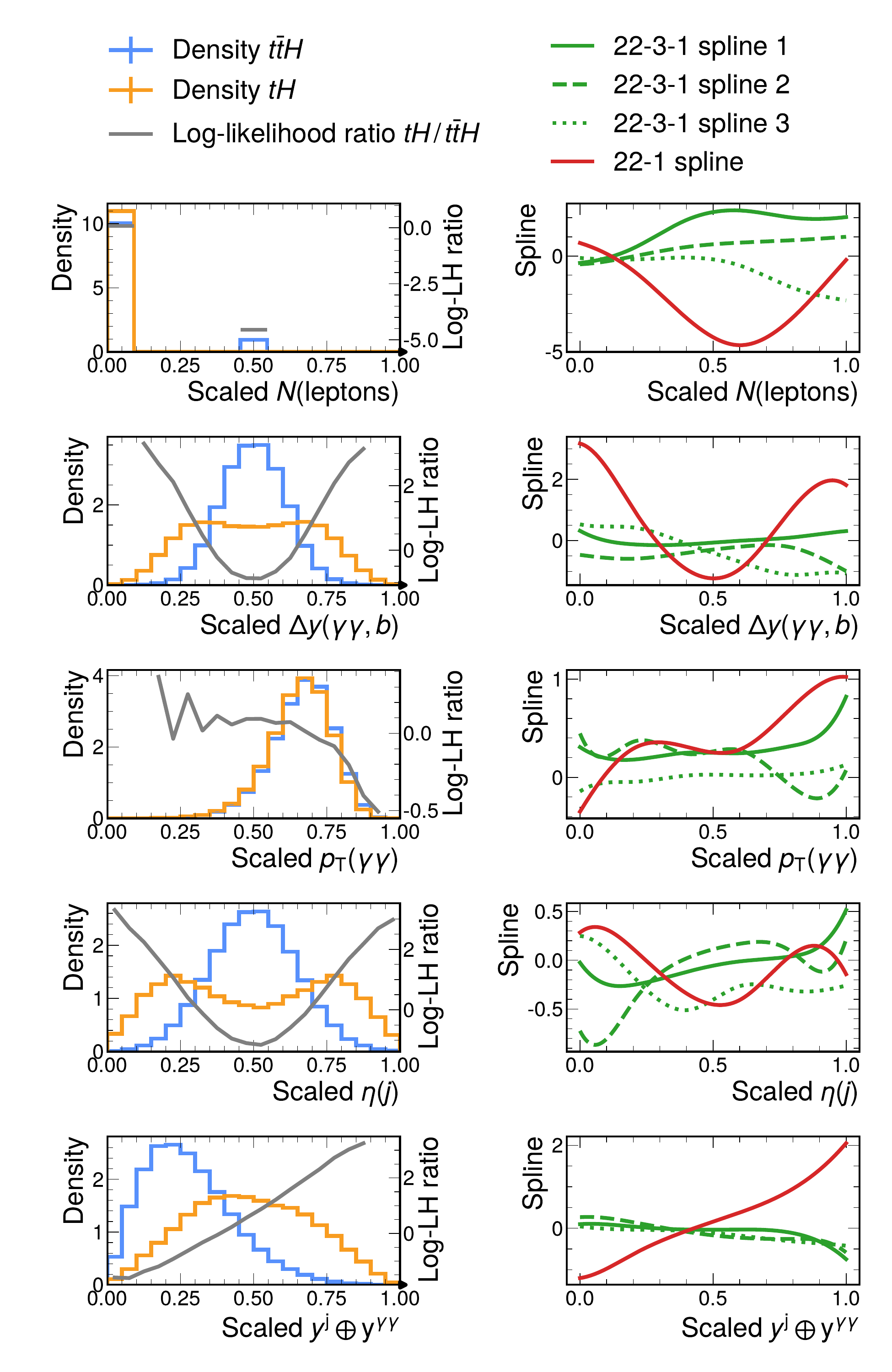}
    \caption{Left: distributions of five examples of pre-processed input features for \tH\ and \ttH\ production together with the corresponding log-likelihood ratio.
    A ratio is shown if there are at least 25 examples of each process from the training dataset in the respective bin.
    Right: the learned spline for these input features in KAN~22--1 (red) and the three learned splines in the first layer of KAN~22--3--1 (green).
    }
    \label{fig:splines_LH_ratio}
\end{figure}

In Fig.~\ref{fig:splines_LH_ratio}, the learned activation functions are shown for five examples of features.
These features were chosen to represent a set of variables with different shapes of their distributions, where the corresponding splines of the single-layer KAN have high $L_1$-norms.
The distributions of the pre-processed features are also shown for the two classes, together with the univariate log-likelihood ratios of the signal over the background.
As expected in binary classification, where the optimal decision boundary is related to the log-likelihood ratio of the classes, the activation functions learned by the one-layer KAN resemble these univariate log-likelihood ratios of the respective input features.
This resemblance is anticipated because the one-layer KAN sums univariate functions of each input feature, aligning with the additive form of the multivariate log-likelihood ratio, which decomposes into a sum of univariate log-likelihood ratios when features are independent.
However, due to correlations, differences between the learned activation functions and the univariate log-likelihood ratios may arise, as we find, for example, in the normalization of the spline transforming the $\eta(j)$ variable.
The corresponding splines obtained from the training of the 22--3--1 KAN are also shown in the same panels.
These differ clearly from the likelihood ratios and the 22--1 KAN splines.
This is consistent with the expectation that multi-layer KANs learn more abstract representations of the data.
As the splines acting on the same input feature are allowed to have different functional shapes or $L_1$-norms, each of the three nodes of the internal layer of the 22--3--1 KAN captures different aspects of the data.

For comparison to the KANs, we train several MLPs with different configurations: 
two shallow MLPs, each with only a single hidden layer of 8 and 32 nodes, respectively, as well as deeper networks with up to five hidden layers.
The largest model has a node count of 22--256--128--64--32--16--1.
The hidden layers are activated by ReLU functions.
The comparison includes models with a number of trainable parameters as small as approximately \num{200} to approximately \num{60000}.

\begin{figure}[p]
    \centering
    \includegraphics[width=0.7\textwidth]{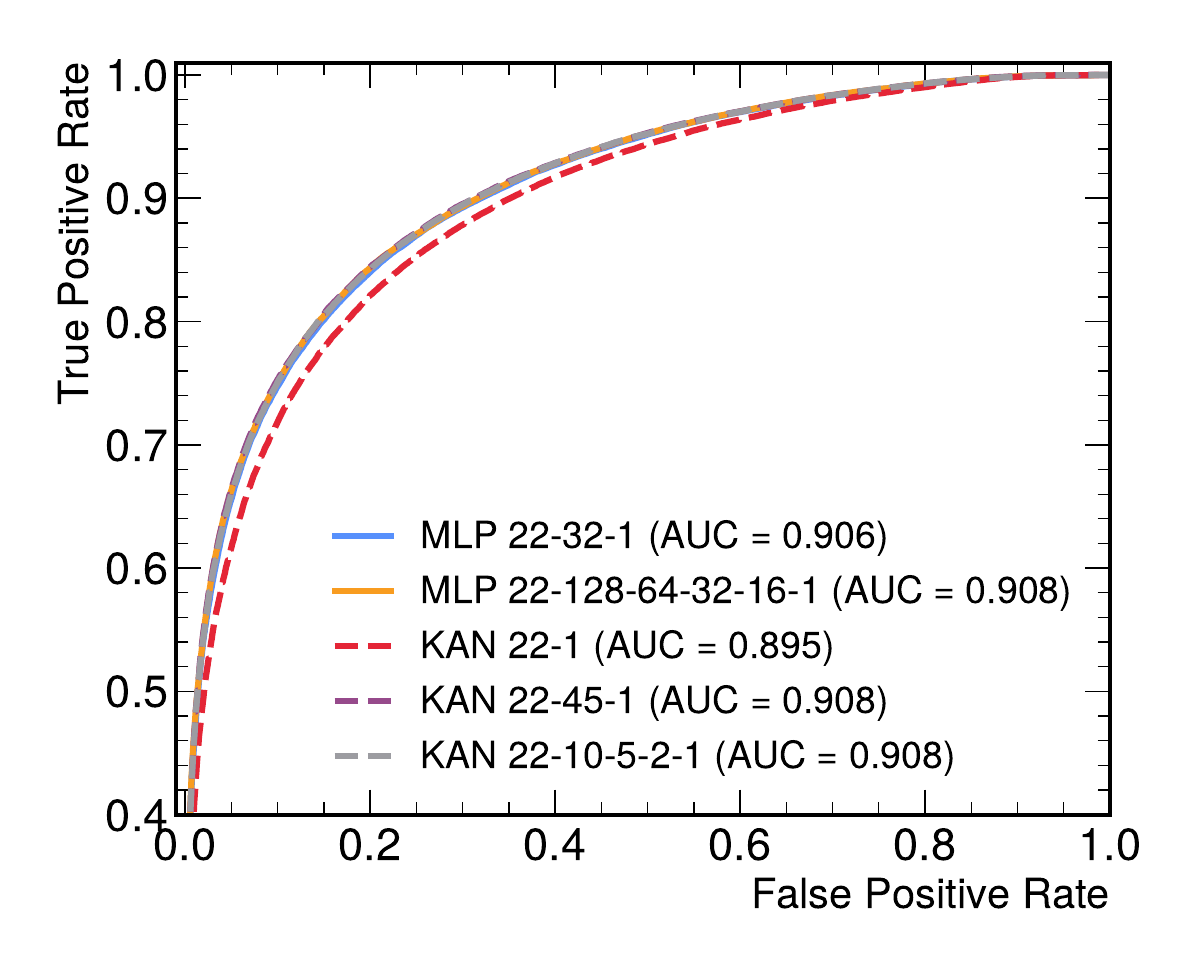}
    \caption{ROC curves of selected models labeled with their node counts and the AUC scores.
    Results from two example MLPs and three example KANs are shown, including deep and shallow models.
    }
    \label{fig:ROC_KAN_MLP}
    \vspace{0.5cm}
    \includegraphics[width=\textwidth]{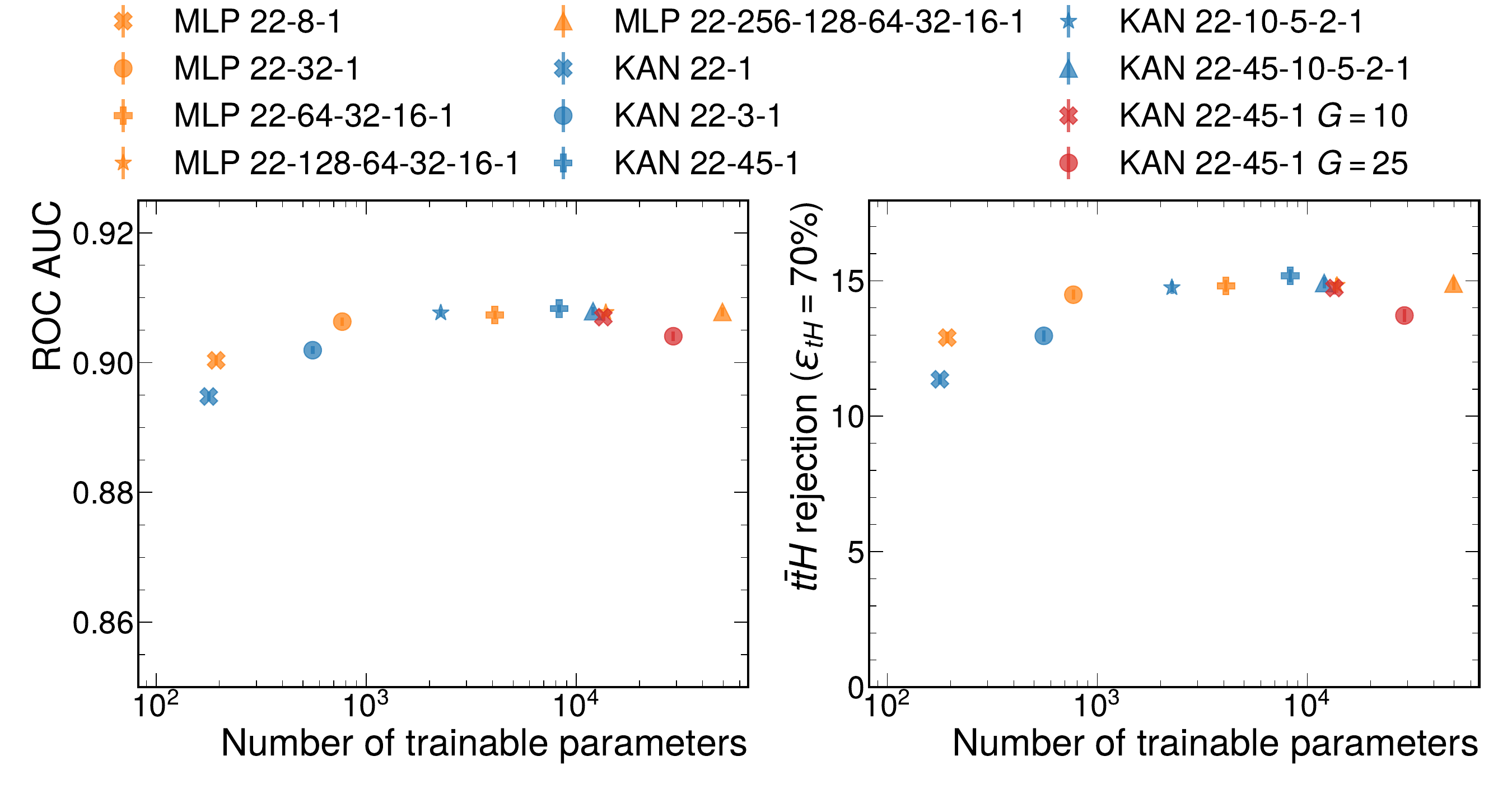}
    \caption{Parameter efficiency of KANs and MLPs: The ROC AUC for different models (left) and the \ttH\ rejection for a \tH\ efficiency of $70\,\%$ (right) are shown as a function of the number of parameters of the different models.
    Error bars corresponding to the limited size of the test dataset are shown, but they are smaller than the symbols.
    }
    \label{fig:AUC_rej_Nparams}
\end{figure}

Receiver operating characteristic (ROC) curves for selected models are shown in Fig.~\ref{fig:ROC_KAN_MLP}, including models with deep and shallow structures for both, MLPs and KANs.
The overall shape of the ROC curves of KANs and MLPs is very similar, except for the one-layer KAN, where the area under the curve (AUC) is considerably lower.
We observe that an MLP with only a single hidden layer reaches an AUC\footnote{The uncertainty in the quoted AUCs from the limited size of the test dataset are approximately $6\times 10^{-4}$.} of $0.906$, only slightly below the AUCs obtained by our best MLP ($0.908$) and the best KAN, where the 22--45--1 network also reaches an AUC of $0.908$.
We find only slight differences in classification performance between well-tuned KANs and MLPs.

To evaluate the parameter efficiency, we compare the performance of KANs and MLPs as a function of the number of trainable model parameters.
A small number of parameters is favorable for a given performance, as smaller models are computationally more efficient, better interpretable (if at all possible), and less prone to overfitting.
In Fig.~\ref{fig:AUC_rej_Nparams}, two metrics often used in HEP for classification performance evaluation are included: the AUC, and the background rejection for a fixed signal efficiency, here chosen as $70\,\%$.
The rejection is defined as the inverse of the efficiency for a given threshold on the network output.
Overall, we find that for very low numbers of parameters, the MLPs outperform the KANs, while for medium and high numbers of parameters, the performance of KANs and MLPs is similar.
The smallest MLP trained with only eight nodes in the hidden layer has only \num{193} parameters and has reached a background rejection of \num{12.9} already\footnote{The uncertainty in the quoted rejection values from the limited size of the test dataset is approximately $0.2$.}.
A similar value is achieved by the 22--3--1 KAN, which has \num{556} trainable parameters.
While increasing the number of parameters of the MLPs, their rejection saturates at around \num{14.9}.
The KANs achieve rejections ranging from \num{11.4} for the single-layer network to \num{15.2} for the 22--45--1 KAN.
The deeper KANs from this study show a similar performance.
Instead of varying the node count of KANs, the number of model parameters can be increased as well by raising the grid parameter $G$.
For illustration, the 22--45--1 KAN is included in Fig.~\ref{fig:AUC_rej_Nparams} when trained with $G=10$ and $G=25$.
For KANs with grid parameters considerably higher than the default of $G=5$, we find that these models overtrain faster and generalize worse.

%% file: tex_inputs/conclusions.tex
We studied the application of Kolmogorov-Arnold Networks (KANs) to a typical binary event classification task in high-energy physics (HEP).
The dataset used contains simulated events of Higgs-boson production in association with a top quark pair and with a single top quark in the \Hgg\ decay channel, where 22 discriminative input features were considered in the network trainings.
We presented studies on the interpretability of KANs, compared their performance and parameter efficiency to traditional multilayer perceptrons (MLPs), and documented our findings in the practical training of KANs.
To our knowledge, this is the first time KANs have been applied in the field of particle physics.

As long as the KANs have a hidden layer with multiple nodes, we observe very similar performance to MLPs.
Numerically, our best KAN is even slightly better than the best MLP, but this difference is very small and most likely practically irrelevant when considering systematic uncertainties in a physics analysis.
These findings seem to contradict Ref.~\cite{KAN}, where a clear improvement over the performance of MLPs was found in several examples.
In these examples, loss values were reached that were orders of magnitude lower than those of MLPs, for example, in fitting $f(x,y) = x \cdot y$.
However, those examples are of much lower dimensionality and complexity than our classification task.
We suppose that the examples in Ref.~\cite{KAN} are especially well-suited for learning the Kolmogorov-Arnold representation of the underlying functional relationship.
However, our dataset includes features with a stronger variability in shape.
The representations that have to be learned to solve our classification task may hence not be particularly suited for the KAN architecture.

We find that MLPs outperform KANs for a very low number of trainable parameters.
However, we note that for our binary classification task, which we consider typical in terms of complexity for event classification at the LHC, using such very small models only comes at a moderate cost regarding performance.
Similar performance of MLPs and KANs is then reached for a number of trainable parameters above approximately \num{1000}.
We conclude that for our task KANs are not more parameter efficient than MLPs.

In terms of interpretability, we find that small KANs indeed offer advantages.
For a one-layer KAN, we observe that the learned activation functions resemble the univariate log-likelihood ratios of the input features which these functions act on.
In addition, the $L_1$-norms of the activation functions offer a straightforward interpretation of the importance of different input features for such KANs with only one layer.
Somewhat larger KANs may still offer an illustrative visualization of the activation functions, which we regard as an intrinsic interpretability advantage of KANs.
However, for KANs of greater depth or with wider layers, interpretability seems challenging.

Because of their better interpretability compared to MLPs, we conclude that KANs are a promising alternative for classification tasks in HEP when the performance of small KANs is sufficient or when moderate performance losses are acceptable in favor of interpretability.
We believe that more research on applying KANs in HEP tasks is necessary.
This study primarily focuses on the performance and parameter efficiency of KANs compared to MLPs, as well as the interpretability of small KANs.
The interpretability of deeper KANs should be explored in future studies to address this limitation of our current work.
Mechanistic interpretability techniques may provide systematic ways to understand the learned representations in these more complex models, and exploring symbolic regression to approximate the learned activation functions could open new avenues for applying KANs as a tool in HEP.
In addition, assessing KAN performance and interpretability in HEP regression tasks may further broaden their applicability beyond classification.

%% file: tex_inputs/acknowledgements.tex
\subsubsection*{Acknowledgements}

This research was supported by the Deutsche Forschungsgemeinschaft (DFG) under grants 400140256 - GRK~2497 (The physics of the heaviest particles at the LHC, all authors) and 686709 - ER~866/1-1 (Heisenberg Programme, JE), and by the Studienstiftung des deutschen Volkes (FM, JLS).

%% file: tex_inputs/appendix.tex
\begin{figure}[h]
    \centering
    \includegraphics[width=\textwidth]{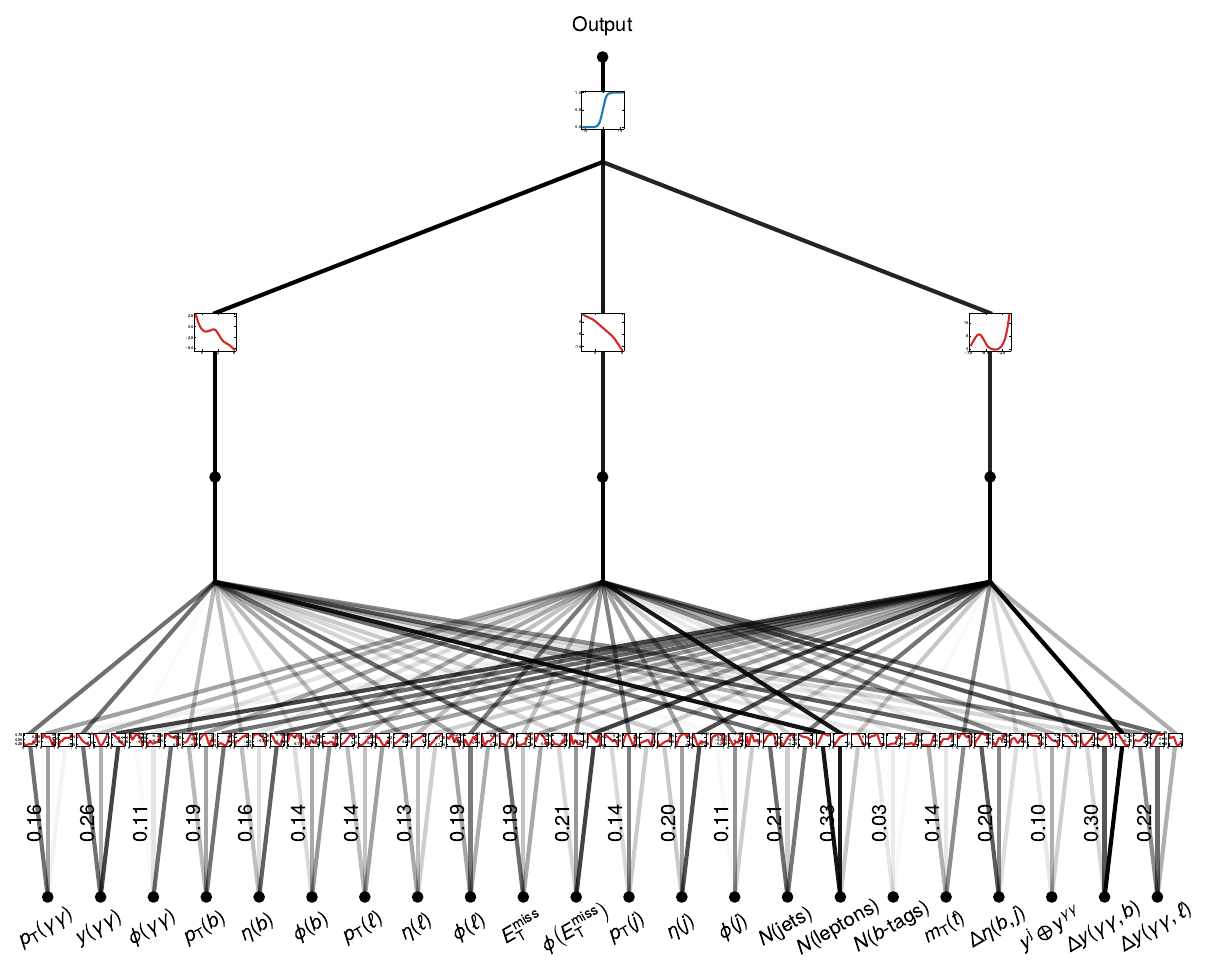}
    \caption{Graphical representation of the trained KAN in the 22--3--1 configuration, i.e. with a second layer with three nodes.
      The red curves represent the learned activation functions, while the blue curve shows the sigmoid function used to normalize the network output.
      The values printed on the edges of the first KAN layer are the $L_1$-norm of each input node, averaged over the three activation functions.}
    \label{fig:App_KAN_22-3-1}
\end{figure}

The graphical representation of the trained KAN~22--3--1 is shown in Fig.~\ref{fig:App_KAN_22-3-1}.
We observe a clear hierarchy in the importance of the different input features for the KAN output, as indicated by the $L_1$-norms on the input nodes.
For the edges with large $L_1$-norms, we observe the tendency towards simple and smooth activation functions.
For edges with lower values of the $L_1$-norms, we also observe more complex activation functions with several local minima and maxima.

%% file: main.bbl
\providecommand{\href}[2]{#2}\begingroup\raggedright\begin{thebibliography}{100}

\bibitem{D0:2009isq}
{D0} collaboration, V.~M. Abazov et~al., \emph{{Observation of Single Top Quark
  Production}},
  \href{http://dx.doi.org/10.1103/PhysRevLett.103.092001}{\emph{Phys. Rev.
  Lett.} {\bfseries 103} (2009) 092001},
  \href{https://arxiv.org/abs/0903.0850}{arXiv:0903.0850}.

\bibitem{CDF:2009itk}
{CDF} collaboration, T.~Aaltonen et~al., \emph{{First Observation of
  Electroweak Single Top Quark Production}},
  \href{http://dx.doi.org/10.1103/PhysRevLett.103.092002}{\emph{Phys. Rev.
  Lett.} {\bfseries 103} (2009) 092002},
  \href{https://arxiv.org/abs/0903.0885}{arXiv:0903.0885}.

\bibitem{Baldi:2014kfa}
P.~Baldi, P.~Sadowski and D.~Whiteson, \emph{{Searching for Exotic Particles in
  High-Energy Physics with Deep Learning}},
  \href{http://dx.doi.org/10.1038/ncomms5308}{\emph{Nature Commun.} {\bfseries
  5} (2014) 4308}, \href{https://arxiv.org/abs/1402.4735}{arXiv:1402.4735}.

\bibitem{Feickert:2021ajf}
M.~Feickert and B.~Nachman, \emph{{A Living Review of Machine Learning for
  Particle Physics}},
  \href{https://arxiv.org/abs/2102.02770}{arXiv:2102.02770}.

\bibitem{Guest:2018yhq}
D.~Guest, K.~Cranmer and D.~Whiteson, \emph{{Deep Learning and its Application
  to LHC Physics}},
  \href{http://dx.doi.org/10.1146/annurev-nucl-101917-021019}{\emph{Ann. Rev.
  Nucl. Part. Sci.} {\bfseries 68} (2018) 161},
  \href{https://arxiv.org/abs/1806.11484}{arXiv:1806.11484}.

\bibitem{Radovic:2018dip}
A.~Radovic, M.~Williams, D.~Rousseau, M.~Kagan, D.~Bonacorsi, A.~Himmel et~al.,
  \emph{{Machine learning at the energy and intensity frontiers of particle
  physics}}, \href{http://dx.doi.org/10.1038/s41586-018-0361-2}{\emph{Nature}
  {\bfseries 560} (2018) 41}.

\bibitem{Schwartz:2021ftp}
M.~D. Schwartz, \emph{{Modern Machine Learning and Particle Physics}},
  \href{http://dx.doi.org/10.1162/99608f92.beeb1183}{\emph{Harvard Data Science
  Review} (2021) 3}, \href{https://arxiv.org/abs/2103.12226}{arXiv:2103.12226}.

\bibitem{Karagiorgi:2021ngt}
G.~Karagiorgi, G.~Kasieczka, S.~Kravitz, B.~Nachman and D.~Shih, \emph{{Machine
  learning in the search for new fundamental physics}},
  \href{http://dx.doi.org/10.1038/s42254-022-00455-1}{\emph{Nature Rev. Phys.}
  {\bfseries 4} (2022) 399}.

\bibitem{interpret_def}
F.~Doshi-Velez and B.~Kim, \emph{{T}owards {A} {R}igorous {S}cience of
  {I}nterpretable {M}achine {L}earning},
  \href{https://arxiv.org/abs/1702.08608}{arXiv:1702.08608}.

\bibitem{XAI}
A.~B. Arrieta, N.~Díaz-Rodríguez, J.~D. Ser, A.~Bennetot, S.~Tabik,
  A.~Barbado et~al., \emph{{Explainable Artificial Intelligence (XAI):
  Concepts, Taxonomies, Opportunities and Challenges toward Responsible AI}},
  \href{https://arxiv.org/abs/1910.10045}{arXiv:1910.10045}.

\bibitem{guidotti2018survey}
R.~Guidotti, A.~Monreale, S.~Ruggieri, F.~Turini, F.~Giannotti and
  D.~Pedreschi, \emph{A survey of methods for explaining black box models},
  {\emph{ACM Computing Surveys} {\bfseries 51} (2018) 1}.

\bibitem{interpretable}
X.~Li, H.~Xiong, X.~Li, X.~Wu, X.~Zhang, J.~Liu et~al., \emph{Interpretable
  deep learning: interpretation, interpretability, trustworthiness, and
  beyond}, \href{http://dx.doi.org/10.1007/s10115-022-01756-8}{\emph{Knowledge
  and Information Systems} {\bfseries 64} (2022) 3197},
  \href{https://arxiv.org/abs/2103.10689}{arXiv:2103.10689}.

\bibitem{SHAP}
S.~Lundberg and S.-I. Lee, \emph{A unified approach to interpreting model
  predictions},  \href{https://arxiv.org/abs/1705.07874}{arXiv:1705.07874}.

\bibitem{Breiman2001}
L.~Breiman, \emph{Random forests},
  \href{http://dx.doi.org/10.1023/A:1010933404324}{\emph{Machine Learning}
  {\bfseries 45} (2001) 5}.

\bibitem{LIME}
M.~T. Ribeiro, S.~Singh and C.~Guestrin, \emph{{"Why Should I Trust You?":
  Explaining the Predictions of Any Classifier}}, {\emph{Proceedings of
  KDD2016} (2016) 1135}.

\bibitem{NAMs}
R.~Agarwal, L.~Melnick, N.~Frosst, X.~Zhang, B.~Lengerich, R.~Caruana et~al.,
  \emph{{Neural Additive Models: Interpretable machine learning with neural
  nets}}, {\emph{Advances in neural information processing systems} {\bfseries
  34} (2021) 4699}.

\bibitem{KAN}
Z.~Liu, Y.~Wang, S.~Vaidya, F.~Ruehle, J.~Halverson, M.~Solja{\v{c}}i{\'c}
  et~al., \emph{{KAN}: Kolmogorov-{A}rnold {N}etworks},
  \href{https://arxiv.org/abs/2404.19756}{arXiv:2404.19756}.

\bibitem{UAT}
K.~Hornik, M.~Stinchcombe and H.~White, \emph{Multilayer feedforward networks
  are universal approximators},
  \href{http://dx.doi.org/10.1016/0893-6080(89)90020-8}{\emph{Neural Networks}
  {\bfseries 2} (1989) 359}.

\bibitem{kolmogorov}
A.~N. Kolmogorov, \emph{On the representation of continuous functions of many
  variables by superposition of continuous functions of one variable and
  addition}, {\emph{Doklady Akademii Nauk} {\bfseries 114} (1957) 953}.

\bibitem{Bohra}
P.~Bohra, J.~Campos, H.~Gupta, S.~Aziznejad and M.~Unser, \emph{{Learning
  Activation Functions in Deep (Spline) Neural Networks}}, {\emph{IEEE Open
  Journal of Signal Processing} {\bfseries 1} (2020) 295}.

\bibitem{Aziznejad}
S.~Aziznejad and M.~Unser, \emph{{Deep Spline Networks with Control of
  Lipschitz Regularity}},  Proceedings of ICASSP 2019, p.~3242.

\bibitem{agostinelli2014learning}
F.~Agostinelli, \emph{Learning activation functions to improve deep neural
  networks},  \href{https://arxiv.org/abs/1412.6830}{arXiv:1412.6830}.

\bibitem{goyal2019learning}
M.~Goyal, R.~Goyal and B.~Lall, \emph{{Learning activation functions: A new
  paradigm for understanding neural networks}},
  \href{https://arxiv.org/abs/1906.09529}{arXiv:1906.09529}.

\bibitem{SNAKE}
L.~Ziyin, T.~Hartwig and M.~Ueda, \emph{Neural networks fail to learn periodic
  functions and how to fix it}, {\emph{Advances in Neural Information
  Processing Systems} {\bfseries 33} (2020) 1583}.

\bibitem{DRA}
B.~Hashemi, R.~G. Corominas and A.~Giacchetto, \emph{Can transformers do
  enumerative geometry?},
  \href{https://arxiv.org/abs/2408.14915}{arXiv:2408.14915}.

\bibitem{zhang2022neural}
S.~Zhang, Z.~Shen and H.~Yang, \emph{Neural network architecture beyond width
  and depth}, {\emph{Advances in Neural Information Processing Systems}
  {\bfseries 35} (2022) 5669}.

\bibitem{sprecher}
D.~A. Sprecher and S.~Draghici, \emph{{Space-filling curves and {K}olmogorov
  superposition-based neural networks}}, {\emph{Neural Networks} {\bfseries 15}
  (2002) 57}.

\bibitem{koeppen}
M.~K{\"o}ppen, \emph{{On the {T}raining of a {K}olmogorov {N}etwork}},
  Proceedings of ICANN 2002, p.~474.

\bibitem{lin}
J.-N. Lin and R.~Unbehauen, \emph{On the {R}ealization of a {K}olmogorov
  {N}etwork}, {\emph{Neural Computation} {\bfseries 5} (1993) 18}.

\bibitem{lai}
M.-J. Lai and Z.~Shen, \emph{The {K}olmogorov {S}uperposition {T}heorem can
  {B}reak the {C}urse of {D}imensionality {W}hen {A}pproximating {H}igh
  {D}imensional {F}unctions},
  \href{https://arxiv.org/abs/2112.09963}{arXiv:2112.09963}.

\bibitem{leni}
P.-E. Leni, Y.~D. Fougerolle and F.~Truchetet, \emph{The {K}olmogorov {S}pline
  {N}etwork for {I}mage {P}rocessing},  Proceedings of Image Processing:
  Concepts, Methodologies, Tools, and Applications, p.~54, 2013.

\bibitem{fakhoury}
D.~Fakhoury, E.~Fakhoury and H.~Speleers, \emph{Ex{S}pli{N}et: An interpretable
  and expressive spline-based neural network}, {\emph{Neural Networks}
  {\bfseries 152} (2022) 332}.

\bibitem{montanelli}
H.~Montanelli and H.~Yang, \emph{Error bounds for deep {R}e{L}{U} networks
  using the {K}olmogorov-{A}rnold superposition theorem}, {\emph{Neural
  Networks} {\bfseries 129} (2020) 1}.

\bibitem{he}
J.~He, \emph{On the {O}ptimal {E}xpressive {P}ower of {R}e{L}{U} {D}{N}{N}s and
  {I}ts {A}pplication in {A}pproximation with {K}olmogorov {S}uperposition
  {T}heorem},  \href{https://arxiv.org/abs/2308.05509}{arXiv:2308.05509}.

\bibitem{duda2024biology}
J.~Duda, \emph{{B}iology-inspired joint distribution neurons based on
  {H}ierarchical {C}orrelation {R}econstruction allowing for multidirectional
  neural networks},  \href{https://arxiv.org/abs/2405.05097}{arXiv:2405.05097}.

\bibitem{li2024kolmogorov}
Z.~Li, \emph{{K}olmogorov-{A}rnold {N}etworks are radial basis function
  networks},  \href{https://arxiv.org/abs/2405.06721}{arXiv:2405.06721}.

\bibitem{genet2024tkan}
R.~Genet and H.~Inzirillo, \emph{{TKAN}: {T}emporal {K}olmogorov-{A}rnold
  {N}etworks},  \href{https://arxiv.org/abs/2405.07344}{arXiv:2405.07344}.

\bibitem{peng2024predictive}
Y.~Peng, M.~He, F.~Hu, Z.~Mao, X.~Huang and J.~Ding, \emph{{Predictive Modeling
  of Flexible EHD Pumps using {K}olmogorov-{A}rnold {N}etworks}},
  \href{https://arxiv.org/abs/2405.07488}{arXiv:2405.07488}.

\bibitem{vaca2024kolmogorov}
C.~J. Vaca-Rubio, L.~Blanco, R.~Pereira and M.~Caus,
  \emph{{K}olmogorov-{A}rnold {N}etworks ({KAN}s) for time series analysis},
  \href{https://arxiv.org/abs/2405.08790}{arXiv:2405.08790}.

\bibitem{samadi2024smooth}
M.~E. Samadi, Y.~M{\"u}ller and A.~Schuppert, \emph{Smooth {K}olmogorov
  {A}rnold {N}etworks enabling structural knowledge representation},
  \href{https://arxiv.org/abs/2405.11318}{arXiv:2405.11318}.

\bibitem{bozorgasl2024wav}
Z.~Bozorgasl and H.~Chen, \emph{{W}av-{KAN}: {W}avelet {K}olmogorov-{A}rnold
  {N}etworks},  \href{https://arxiv.org/abs/2405.12832}{arXiv:2405.12832}.

\bibitem{yang2024endowing}
S.~Yang, L.~Qin and X.~Yu, \emph{{E}ndowing {I}nterpretability for {N}eural
  {C}ognitive {D}iagnosis by {E}fficient {K}olmogorov-{A}rnold {N}etworks},
  \href{https://arxiv.org/abs/2405.14399}{arXiv:2405.14399}.

\bibitem{abueidda2024deepokan}
D.~W. Abueidda, P.~Pantidis and M.~E. Mobasher, \emph{Deep{OKAN}: {D}eep
  {O}perator {N}etwork {B}ased on {K}olmogorov {A}rnold {N}etworks for
  mechanics problems},
  \href{https://arxiv.org/abs/2405.19143}{arXiv:2405.19143}.

\bibitem{cheon2024kolmogorov}
M.~Cheon, \emph{{K}olmogorov-{A}rnold {N}etwork for {S}atellite {I}mage
  {C}lassification in {R}emote {S}ensing},
  \href{https://arxiv.org/abs/2406.00600}{arXiv:2406.00600}.

\bibitem{xu2024fourierkan}
J.~Xu, Z.~Chen, J.~Li, S.~Yang, W.~Wang, X.~Hu et~al.,
  \emph{Fourier{KAN}-{GCF}: {F}ourier {K}olmogorov-{A}rnold {N}etwork--{A}n
  {E}ffective and {E}fficient {F}eature {T}ransformation for {G}raph
  {C}ollaborative {F}iltering},
  \href{https://arxiv.org/abs/2406.01034}{arXiv:2406.01034}.

\bibitem{xu2024kolmogorov}
K.~Xu, L.~Chen and S.~Wang, \emph{{K}olmogorov-{A}rnold {N}etworks for {T}ime
  {S}eries: {B}ridging {P}redictive {P}ower and {I}nterpretability},
  \href{https://arxiv.org/abs/2406.02496}{arXiv:2406.02496}.

\bibitem{genet2024temporal}
R.~Genet and H.~Inzirillo, \emph{{A} {T}emporal {K}olmogorov-{A}rnold
  {T}ransformer for {T}ime {S}eries {F}orecasting},
  \href{https://arxiv.org/abs/2406.02486}{arXiv:2406.02486}.

\bibitem{nehma2024leveraging}
G.~Nehma and M.~Tiwari, \emph{Leveraging {KAN}s {F}or {E}nhanced {D}eep
  {K}oopman {O}perator {D}iscovery},
  \href{https://arxiv.org/abs/2406.02875}{arXiv:2406.02875}.

\bibitem{li2024u}
C.~Li, X.~Liu, W.~Li, C.~Wang, H.~Liu and Y.~Yuan, \emph{{U-KAN Makes Strong
  Backbone for Medical Image Segmentation and Generation}},
  \href{https://arxiv.org/abs/2406.02918}{arXiv:2406.02918}.

\bibitem{shukla2024comprehensive}
K.~Shukla, J.~D. Toscano, Z.~Wang, Z.~Zou and G.~E. Karniadakis, \emph{{A
  comprehensive and FAIR comparison between MLP and KAN representations for
  differential equations and operator networks}},
  \href{https://arxiv.org/abs/2406.02917}{arXiv:2406.02917}.

\bibitem{herbozo2024kan}
L.~F. Herbozo~Contreras, J.~Cui, L.~Yu, Z.~Huang, A.~Nikpour and O.~Kavehei,
  \emph{{KAN-EEG: Towards Replacing Backbone-MLP for an Effective Seizure
  Detection System}},
  \href{https://arxiv.org/abs/medRxiv:10.1101/2024.06.05.24308471}{arXiv:medRxiv:10.1101/2024.06.05.24308471}.

\bibitem{kiamari2024gkan}
M.~Kiamari, M.~Kiamari and B.~Krishnamachari, \emph{{GKAN: Graph
  Kolmogorov-Arnold Networks}},
  \href{https://arxiv.org/abs/2406.06470}{arXiv:2406.06470}.

\bibitem{aghaei2024fkan}
A.~A. Aghaei, \emph{{fKAN: Fractional {K}olmogorov-{A}rnold {N}etworks with
  trainable Jacobi basis functions}},
  \href{https://arxiv.org/abs/2406.07456}{arXiv:2406.07456}.

\bibitem{seydi2024unveiling}
S.~T. Seydi, \emph{{Unveiling the Power of Wavelets: A Wavelet-based
  {K}olmogorov-{A}rnold {N}etwork for Hyperspectral Image Classification}},
  \href{https://arxiv.org/abs/2406.07869}{arXiv:2406.07869}.

\bibitem{azam2024suitability}
B.~Azam and N.~Akhtar, \emph{{Suitability of {KAN}s for Computer Vision: A
  preliminary investigation}},
  \href{https://arxiv.org/abs/2406.09087}{arXiv:2406.09087}.

\bibitem{chen2024sckansformer}
Y.~Chen, Z.~Zhu, S.~Zhu, L.~Qiu, B.~Zou, F.~Jia et~al., \emph{{SCKansformer:
  Fine-Grained Classification of Bone Marrow Cells via Kansformer Backbone and
  Hierarchical Attention Mechanisms}},
  \href{https://arxiv.org/abs/2406.09931}{arXiv:2406.09931}.

\bibitem{wang2024kolmogorov}
Y.~Wang, J.~Sun, J.~Bai, C.~Anitescu, M.~S. Eshaghi, X.~Zhuang et~al.,
  \emph{{Kolmogorov Arnold Informed neural network: A physics-informed deep
  learning framework for solving PDEs based on {K}olmogorov {A}rnold
  {N}etworks}},  \href{https://arxiv.org/abs/2406.11045}{arXiv:2406.11045}.

\bibitem{ta2024bsrbf}
H.-T. Ta, \emph{{BSRBF-KAN: A combination of B-splines and Radial Basic
  Functions in {K}olmogorov-{A}rnold {N}etworks}},
  \href{https://arxiv.org/abs/2406.11173}{arXiv:2406.11173}.

\bibitem{zhang2024graphkan}
F.~Zhang and X.~Zhang, \emph{{GraphKAN: Enhancing Feature Extraction with Graph
  {K}olmogorov {A}rnold {N}etworks}},
  \href{https://arxiv.org/abs/2406.13597}{arXiv:2406.13597}.

\bibitem{bodner2024convolutional}
A.~D. Bodner, A.~S. Tepsich, J.~N. Spolski and S.~Pourteau, \emph{Convolutional
  {K}olmogorov-{A}rnold {N}etworks},
  \href{https://arxiv.org/abs/2406.13155}{arXiv:2406.13155}.

\bibitem{poeta2024benchmarking}
E.~Poeta, F.~Giobergia, E.~Pastor, T.~Cerquitelli and E.~Baralis, \emph{{A
  Benchmarking Study of {K}olmogorov-{A}rnold {N}etworks on Tabular Data}},
  \href{https://arxiv.org/abs/2406.14529}{arXiv:2406.14529}.

\bibitem{aghaei2024rkan}
A.~A. Aghaei, \emph{{rKAN: Rational {K}olmogorov-{A}rnold {N}etworks}},
  \href{https://arxiv.org/abs/2406.14495}{arXiv:2406.14495}.

\bibitem{cheon2024demonstrating}
M.~Cheon, \emph{{Demonstrating the Efficacy of {K}olmogorov-{A}rnold {N}etworks
  in Vision Tasks}},
  \href{https://arxiv.org/abs/2406.14916}{arXiv:2406.14916}.

\bibitem{de2024kolmogorov}
G.~De~Carlo, A.~Mastropietro and A.~Anagnostopoulos, \emph{{Kolmogorov-Arnold
  Graph Neural Networks}},
  \href{https://arxiv.org/abs/2406.18354}{arXiv:2406.18354}.

\bibitem{bresson2024kagnns}
R.~Bresson, G.~Nikolentzos, G.~Panagopoulos, M.~Chatzianastasis, J.~Pang and
  M.~Vazirgiannis, \emph{{KAGNNs: {K}olmogorov-{A}rnold {N}etworks meet Graph
  Learning}},  \href{https://arxiv.org/abs/2406.18380}{arXiv:2406.18380}.

\bibitem{howard2024finite}
A.~A. Howard, B.~Jacob, S.~H. Murphy, A.~Heinlein and P.~Stinis, \emph{Finite
  basis {K}olmogorov-{A}rnold {N}etworks: domain decomposition for data-driven
  and physics-informed problems},
  \href{https://arxiv.org/abs/2406.19662}{arXiv:2406.19662}.

\bibitem{wang2024spectralkan}
Y.~Wang, X.~Yu, Y.~Gao, J.~Sha, J.~Wang, L.~Gao et~al., \emph{{SpectralKAN:
  {K}olmogorov-{A}rnold {N}etwork for Hyperspectral Images Change Detection}},
  \href{https://arxiv.org/abs/2407.00949}{arXiv:2407.00949}.

\bibitem{lobanov2024hyperkan}
V.~Lobanov, N.~Firsov, E.~Myasnikov, R.~Khabibullin and A.~Nikonorov,
  \emph{{HyperKAN: {K}olmogorov-{A}rnold {N}etworks make Hyperspectral Image
  Classificators Smarter}},
  \href{https://arxiv.org/abs/2407.05278}{arXiv:2407.05278}.

\bibitem{dong2024tckin}
F.~Dong, \emph{{TCKIN: A Novel Integrated Network Model for Predicting
  Mortality Risk in Sepsis Patients}},
  \href{https://arxiv.org/abs/2407.06560}{arXiv:2407.06560}.

\bibitem{lawan2024mambaforgcn}
A.~Lawan, J.~Pu, H.~Yunusa, A.~Umar and M.~Lawan, \emph{{MambaForGCN: Enhancing
  Long-Range Dependency with State Space Model and {K}olmogorov-{A}rnold
  {N}etworks for Aspect-Based Sentiment Analysis}},
  \href{https://arxiv.org/abs/2407.10347}{arXiv:2407.10347}.

\bibitem{altarabichi2024dropkan}
M.~G. Altarabichi, \emph{{DropKAN: Regularizing KANs by masking
  post-activations}},
  \href{https://arxiv.org/abs/2407.13044}{arXiv:2407.13044}.

\bibitem{shen2024reduced}
H.~Shen, C.~Zeng, J.~Wang and Q.~Wang, \emph{{Reduced Effectiveness of
  {K}olmogorov-{A}rnold {N}etworks on Functions with Noise}},
  \href{https://arxiv.org/abs/2407.14882}{arXiv:2407.14882}.

\bibitem{inzirillo2024deep}
H.~Inzirillo, \emph{{Deep State Space Recurrent Neural Networks for Time Series
  Forecasting}},  \href{https://arxiv.org/abs/2407.15236}{arXiv:2407.15236}.

\bibitem{troy2024sparks}
W.~Troy, \emph{{Sparks of Quantum Advantage and Rapid Retraining in Machine
  Learning}},  \href{https://arxiv.org/abs/2407.16020}{arXiv:2407.16020}.

\bibitem{toscano2024inferring}
J.~D. Toscano, T.~K{\"a}ufer, M.~Maxey, C.~Cierpka and G.~E. Karniadakis,
  \emph{{Inferring turbulent velocity and temperature fields and their
  statistics from Lagrangian velocity measurements using physics-informed
  {K}olmogorov-{A}rnold {N}etworks}},
  \href{https://arxiv.org/abs/2407.15727}{arXiv:2407.15727}.

\bibitem{li2024coeff}
X.~Li, Z.~Feng, Y.~Chen, W.~Dai, Z.~He, Y.~Zhou et~al., \emph{{COEFF-KANs: A
  Paradigm to Address the Electrolyte Field with KANs}},
  \href{https://arxiv.org/abs/2407.20265}{arXiv:2407.20265}.

\bibitem{rigas2024adaptive}
S.~Rigas, M.~Papachristou, T.~Papadopoulos, F.~Anagnostopoulos and
  G.~Alexandridis, \emph{{Adaptive Training of Grid-Dependent Physics-Informed
  {K}olmogorov-{A}rnold {N}etworks}},
  \href{https://arxiv.org/abs/2407.17611}{arXiv:2407.17611}.

\bibitem{le2024exploring}
T.~X.~H. Le, T.~D. Tran, H.~L. Pham, V.~T.~D. Le, T.~H. Vu, V.~T. Nguyen
  et~al., \emph{{Exploring the Limitations of {K}olmogorov-{A}rnold {N}etworks
  in Classification: Insights to Software Training and Hardware
  Implementation}},  \href{https://arxiv.org/abs/2407.17790}{arXiv:2407.17790}.

\bibitem{seguelvlp}
F.~Seguel, D.~Salihu, S.~H{\"a}gele and E.~Steinbach, \emph{{VLP-KAN:
  Low-complexity and Interpretable RSS-based Visible Light Positioning using
  {K}olmogorov-{A}rnold {N}etworks}},  2024.

\bibitem{zeydan2024f}
E.~Zeydan, C.~J. Vaca-Rubio, L.~Blanco, R.~Pereira, M.~Caus and A.~Aydeger,
  \emph{{F-KANs: Federated {K}olmogorov-{A}rnold {N}etworks}},
  \href{https://arxiv.org/abs/2407.20100}{arXiv:2407.20100}.

\bibitem{zinage2024dkl}
S.~Zinage, S.~Mondal and S.~Sarkar, \emph{{DKL-KAN: Scalable Deep Kernel
  Learning using {K}olmogorov-{A}rnold {N}etworks}},
  \href{https://arxiv.org/abs/2407.21176}{arXiv:2407.21176}.

\bibitem{pratyush2024calmphoskan}
P.~Pratyush, C.~Carrier, S.~Pokharel, H.~D. Ismail, M.~Chaudhari and D.~B. KC,
  \emph{{CaLMPhosKAN: Prediction of General Phosphorylation Sites in Proteins
  via Fusion of Codon Aware Embeddings with Amino Acid Aware Embeddings and
  Wavelet-based {K}olmogorov {A}rnold {N}etwork}},
  \href{https://arxiv.org/abs/bioRxiv:10.1101/2024.07.30.605530}{arXiv:bioRxiv:10.1101/2024.07.30.605530}.

\bibitem{altarabichi2024rethinking}
M.~G. Altarabichi, \emph{{Rethinking the Function of Neurons in KANs}},
  \href{https://arxiv.org/abs/2407.20667}{arXiv:2407.20667}.

\bibitem{liu2024complexity}
H.~Liu, J.~Lei and Z.~Ren, \emph{{From Complexity to Clarity:
  {K}olmogorov-{A}rnold {N}etworks in Nuclear Binding Energy Prediction}},
  \href{https://arxiv.org/abs/2407.20737}{arXiv:2407.20737}.

\bibitem{tang20243d}
T.~Tang, Y.~Chen and H.~Shu, \emph{{3D U-KAN Implementation for Multi-modal MRI
  Brain Tumor Segmentation}},
  \href{https://arxiv.org/abs/2408.00273}{arXiv:2408.00273}.

\bibitem{li2024gnnmolkanharnessingpowerkan}
R.~Li, \emph{{GNN-MolKAN: Harnessing the Power of KAN to Advance Molecular
  Representation Learning with GNNs}},
  \href{https://arxiv.org/abs/2408.01018}{arXiv:2408.01018}.

\bibitem{Farina:2012xp}
M.~Farina, C.~Grojean, F.~Maltoni, E.~Salvioni and A.~Thamm, \emph{{Lifting
  degeneracies in Higgs couplings using single top production in association
  with a Higgs boson}},
  \href{http://dx.doi.org/10.1007/JHEP05(2013)022}{\emph{JHEP} {\bfseries 05}
  (2013) 022}, \href{https://arxiv.org/abs/1211.3736}{arXiv:1211.3736}.

\bibitem{Bahl:2020wee}
H.~Bahl, P.~Bechtle, S.~Heinemeyer, J.~Katzy, T.~Klingl, K.~Peters et~al.,
  \emph{{Indirect $\mathcal{CP}$ probes of the Higgs-top-quark interaction:
  current LHC constraints and future opportunities}},
  \href{http://dx.doi.org/10.1007/JHEP11(2020)127}{\emph{JHEP} {\bfseries 11}
  (2020) 127}, \href{https://arxiv.org/abs/2007.08542}{arXiv:2007.08542}.

\bibitem{CMS:2018uxb}
{CMS} collaboration, A.~M. Sirunyan et~al., \emph{{Observation of
  $\mathrm{t\overline{t}H}$ production}},
  \href{http://dx.doi.org/10.1103/PhysRevLett.120.231801}{\emph{Phys. Rev.
  Lett.} {\bfseries 120} (2018) 231801},
  \href{https://arxiv.org/abs/1804.02610}{arXiv:1804.02610}.

\bibitem{ATLAS:2018mme}
{ATLAS} collaboration, M.~Aaboud et~al., \emph{{Observation of Higgs boson
  production in association with a top quark pair at the LHC with the ATLAS
  detector}},
  \href{http://dx.doi.org/10.1016/j.physletb.2018.07.035}{\emph{Phys. Lett. B}
  {\bfseries 784} (2018) 173},
  \href{https://arxiv.org/abs/1806.00425}{arXiv:1806.00425}.

\bibitem{CMS:2024fdo}
{CMS} collaboration, A.~Hayrapetyan et~al., \emph{{Measurement of the
  $\mathrm{t\bar{t}H}$ and $\mathrm{tH}$ production rates in the
  $\mathrm{H}\to\mathrm{b\bar{b}}$ decay channel using proton-proton collision
  data at $\sqrt{s} = 13\,\mathrm{TeV}$}},
  \href{https://arxiv.org/abs/2407.10896}{arXiv:2407.10896}.

\bibitem{ATLAS:2022vkf}
{ATLAS} collaboration, G.~Aad et~al., \emph{{A detailed map of Higgs boson
  interactions by the ATLAS experiment ten years after the discovery}},
  \href{http://dx.doi.org/10.1038/s41586-022-04893-w}{\emph{Nature} {\bfseries
  607} (2022) 52}, \href{https://arxiv.org/abs/2207.00092}{arXiv:2207.00092}.

\bibitem{AMCAT}
J.~Alwall, R.~Frederix, S.~Frixione, V.~Hirschi, F.~Maltoni, O.~Mattelaer
  et~al., \emph{The automated computation of tree-level and next-to-leading
  order differential cross sections, and their matching to parton shower
  simulations}, \href{http://dx.doi.org/10.1007/JHEP07(2014)079}{\emph{JHEP}
  {\bfseries 07} (2014) 079},
  \href{https://arxiv.org/abs/1405.0301}{arXiv:1405.0301}.

\bibitem{Ball:2012cx}
R.~D. Ball et~al., \emph{{Parton distributions with LHC data}},
  \href{http://dx.doi.org/10.1016/j.nuclphysb.2012.10.003}{\emph{Nucl. Phys. B}
  {\bfseries 867} (2013) 244},
  \href{https://arxiv.org/abs/1207.1303}{arXiv:1207.1303}.

\bibitem{Demartin:2015uha}
F.~Demartin, F.~Maltoni, K.~Mawatari and M.~Zaro, \emph{{Higgs production in
  association with a single top quark at the LHC}},
  \href{http://dx.doi.org/10.1140/epjc/s10052-015-3475-9}{\emph{Eur. Phys. J.
  C} {\bfseries 75} (2015) 267},
  \href{https://arxiv.org/abs/1504.00611}{arXiv:1504.00611}.

\bibitem{loop_induced}
V.~Hirschi and O.~Mattelaer, \emph{{Automated event generation for loop-induced
  processes}}, \href{http://dx.doi.org/10.1007/JHEP10(2015)146}{\emph{JHEP}
  {\bfseries 10} (2015) 146},
  \href{https://arxiv.org/abs/1507.00020}{arXiv:1507.00020}.

\bibitem{Bierlich:2022pfr}
C.~Bierlich et~al., \emph{{A comprehensive guide to the physics and usage of
  PYTHIA 8.3}},
  \href{http://dx.doi.org/10.21468/SciPostPhysCodeb.8}{\emph{SciPost Phys.
  Codeb.} (2022) 8}, \href{https://arxiv.org/abs/2203.11601}{arXiv:2203.11601}.

\bibitem{deFavereau:2013fsa}
{DELPHES 3} collaboration, J.~de~Favereau, C.~Delaere, P.~Demin, A.~Giammanco,
  V.~Lema\^\i{}tre, A.~Mertens et~al., \emph{{DELPHES 3: a modular framework
  for fast simulation of a generic collider experiment}},
  \href{http://dx.doi.org/10.1007/JHEP02(2014)057}{\emph{JHEP} {\bfseries 02}
  (2014) 057}, \href{https://arxiv.org/abs/1307.6346}{arXiv:1307.6346}.

\bibitem{Cacciari:2008gp}
M.~Cacciari, G.~P. Salam and G.~Soyez, \emph{{The anti-$k_t$ jet clustering
  algorithm}},
  \href{http://dx.doi.org/10.1088/1126-6708/2008/04/063}{\emph{JHEP} {\bfseries
  04} (2008) 063}, \href{https://arxiv.org/abs/0802.1189}{arXiv:0802.1189}.

\bibitem{Cacciari:2011ma}
M.~Cacciari, G.~P. Salam and G.~Soyez, \emph{{FastJet User Manual}},
  \href{http://dx.doi.org/10.1140/epjc/s10052-012-1896-2}{\emph{Eur. Phys. J.
  C} {\bfseries 72} (2012) 1896},
  \href{https://arxiv.org/abs/1111.6097}{arXiv:1111.6097}.

\bibitem{paszke2019pytorch}
A.~Paszke, S.~Gross, F.~Massa, A.~Lerer, J.~Bradbury, G.~Chanan et~al.,
  \emph{Pytorch: An imperative style, high-performance deep learning library},
  Proceedings of NeurIPS 2019.
\newblock \href{https://arxiv.org/abs/1912.01703}{arXiv:1912.01703}.

\bibitem{tensorflow_2_17_0}
{TensorFlow Developers}, \emph{{TensorFlow v2.17.0}},  2024.
\newblock 10.5281/zenodo.12726004.

\bibitem{kingma2017adam}
D.~P. Kingma and J.~Ba, \emph{{Adam: A Method for Stochastic Optimization}},
  Proceedings of ICLR 2015.
\newblock \href{https://arxiv.org/abs/1412.6980}{arXiv:1412.6980}.

\bibitem{LBFGS}
D.~C. Liu and J.~Nocedal, \emph{{On the limited memory BFGS method for large
  scale optimization}},
  \href{http://dx.doi.org/10.1007/BF01589116}{\emph{Mathematical Programming}
  {\bfseries 45} (1989) 503}.

\end{thebibliography}\endgroup
